\documentclass[%
 reprint,
showpacs,
 amsmath,amssymb,
 aps,
prb,
]{revtex4-1}

\usepackage{graphicx}
\usepackage{dcolumn}
\usepackage{bm}
\usepackage{hyperref}
\usepackage{mathrsfs}
\usepackage{subfigure}
\usepackage{float}
\usepackage{SIunits}
\newenvironment{shrinkeq}[1]
{ \bgroup
  \addtolength\abovedisplayshortskip{#1}
  \addtolength\abovedisplayskip{#1}
  \addtolength\belowdisplayshortskip{#1}
  \addtolength\belowdisplayskip{#1}}
{\egroup\ignorespacesafterend}
\begin{document}

\preprint{prb/123-QED}

\title{Majorana Fermions in Periodically Driven Semiconductor-Superconductor Heterostructure}
\author{Chen-Cheng Wu$^{1,2}$}
\author{Jiao Sun$^{1,2}$}
\author{Fei-Jie Huang$^{1}$}%
\author{Yun-De Li$^{2}$}
\author{Wu-Ming Liu$^{1}$}
\affiliation{%
 $^1$Beijing National Laboratory for Condensed Matter Physics, Institute of Physics, Chinese Academy of Science, Beijing 100190, China
}%
\affiliation{%
 $^2$School of Physical Science and Technology, Yunnan University, Kunming 650000, China
}%




\date{\today}

\begin{abstract}
We propose a new approach to create  Majorana fermions at the edge of a periodically driven semiconductor-superconductor Heterostructure.\ We calculate the quasi-energy spectrum of the periodically driven Heterostructure by using the Floquet's theory.\ When the interaction between different Brillouin zones of quasi-energy is neglected, one Majorana fermion can be created at each edge of the Heterostructure when the ratio of driven amplitude and driven frequency is larger than a minimum.\ Furthermore, when the interaction between the nearest Brillouin zones of quasi-energy is considered, we restrict the condition of creating Majorana fermions above with a lower limit of the driven frequency.\ We also discuss the experimental protocol of creating Majorana fermions in the periodically driven Heterostructure.
\end{abstract}
\maketitle
\section{\label{sec1}INTRODUCTION}
Majorana fermions are so unique because a Majorana fermion is its own antiparticle.\cite{Wilczek2009,Franz2010,Stern2010} $2n$ well separated Majorana bound states can construct $n$ ordinary fermions.\ Because a Majorana fermion is its own antiparticle, Majorana fermions can be excited without energy, which makes the ground state degenerate, quantum information can be encoded into these degenerate states, which is protected from the decoherence.\cite{Kitaev2003} Braiding Majorana fermions around one another transform the state to the other degenerate states,\cite{Read2000,Ivanov2001,Alicea2011} thus quantum information encoded in this degenerate state can be manipulated through these transforming.\ Therefore, the Majorana fermions has great potential for topological quantum computation.\cite{Kitaev2003,Nayak2008,Hassler2010,Sau2010,Bonderson2011,Flensberg2011}\\
\indent It is predicted that 1D Kitaev model would host one unpaired Majorana fermion at each edge under topological phase transition.\cite{Kitaev2001} Recently, several experimental protocols are proposed to realize Majorana fermions in static systems.\cite{Lutchyn2010,jiang2011,Niu2011,shouchengzhang2009,shouchengzhang2009B,xiaoliangqi2010,xiaoliangqi2011,karyn2009,Sarma2005,Fu2008,Volovik2009,Linder2010,Wimmer2010,Potter2010,Flensberg2010,Akhmerov2011,Fulga2011,Brouwer20112,Potter2011,Potter20112,Brouwer2011} On the other hand, Majorana fermions have also been studied in time-dependent systems.\ By applying Floquet's theory, it has been shown that time-dependent systems can develop topological phases that have no analog of static systems.\cite{Kitagawa2010,Lindner2010,Kitagawa2012,Kitagawa2010A} Majorana fermions have been predicted in cold-atom quantum wire (with static spin-orbit coupling and magnetic field) which is driven by an effective time-periodic chemical potential.\cite{jiang2011}  Meanwhile, Majorana fermions were predicted to appear at the edge of a 2D cold-atom superfluid system in which the potential of the optical square lattice is periodically varied.\cite{liu2012} In addition, Majorana fermions have been predicted in a 1D driven Heterostructure without magnetic fields.\cite{Reynoso2013} Although there have been many experimental protocols of creating Majorana fermions in static and driven system, and some of them have been studied in real experiments,\cite{Das2012,Mourik2012} it still don't have enough evidence of the existence of the Majorana fermions in such experiments.\ Furthermore, a semiconductor-superconductor Heterostructure with a periodically driven chemical potential hasn't been studied before, so it is interesting to find a new way of creating Majorana fermions in such system.\\
\begin{shrinkeq}{-2ex}
\begin{figure}
\includegraphics[scale=0.35]{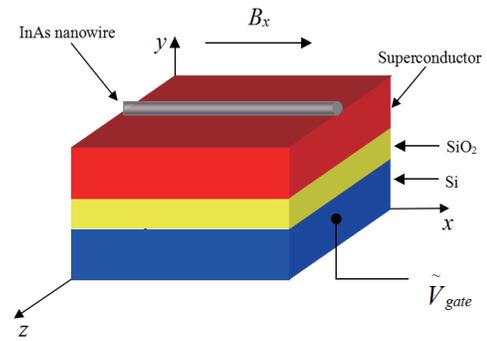}
\caption{Schematic of periodically driven semiconductor-superconductor Heterostructure.\ InAs nanowire is contacted with an ordinary superconductor.\ The Cooper pairs of the superconductor leak into semiconductor via the proximity effect.\ The superconductor is separated from the Si substrate by a SiO$_2$ layer.\ An alternating gatevoltage $\widetilde{V}_{gate}$ is applied to the Si substrate to vary the electron density of the nanowire, which changes the chemical potential of the nanowire periodically.\ The magnetic field $B_x$ is applied along $x$ direction. }
\label{fig5}
\end{figure}
\end{shrinkeq}
\indent In this paper, we propose a new approach to create Majorana fermions in a periodically driven semiconductor-superconductor Heterostructure.\ We first discuss a static Heterostructure in which spin-orbit coupling, $s$-wave pairing field and Zeeman interaction coexist.\ A system which contains effects above has been studied much before.\cite{jiang2011,Lutchyn2010,karyn2010,liu2005,liu2012B,Niu2013} On this basis, a periodically driven chemical potential is applied to the Heterostructure.\ Utilzing Floquet's theory, we demonstrate that Majorana fermions can be created under certain conditions, where the ratio of driven amplitude and driven frequency should be larger than a minimum and the driven frequency has a lower limit.\\
\indent This paper is organized as follows. In Sec.\ \ref{sec2}, we introduce the periodically driven Heterostructure and point out the condition of creating Majorana fermions in static Heterostructure.\ In Sec.\ \ref{sec3}, we first introduce the Floquet's theory briefly.\ Then we use Floquet's theory to calculate the quasi-energy spectrums of the Heterostructure with some approximations.\ Here we discuss the case of neglecting the interaction between different Brillouin zones of quasi-energy first, in this case, we can just consider the diagonal elements of the Floquet Hamiltonian and ignore the other part of the Hamiltonian.\ Then we discuss the case of considering the nearest Brillouin zones of quasi-energy, this can be done by adding considering the elements near the diagonal elements.\ With the help of the discussions above, we find the conditions of creating Majorana fermions in our Heterostructure.\ That is, for certain values of cooper pairing field and Zeeman interaction, the ratio of driven amplitude and driven frequency should be larger than a minimum, furthermore, the driven frequency has a lower limit.\ In Sec.\ \ref{sec4}, we discuss the experimental protocol briefly.
\section{\label{sec2}PERIODICALLY DRIVEN HETEROSTRUCTURE}
\begin{figure}
\includegraphics[scale=0.38]{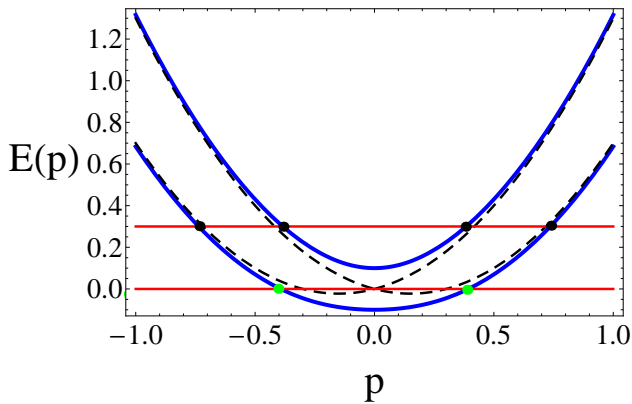}
\caption{Energy spectrums $E(p)$ of $\mathcal{H}_0$ in Eq.\ \eqref{equ2} when pairing field $\Delta=0$, the spin-orbit coupling strength $\alpha=0.3$ (setting $1/2m$ to be unit).\ The black dashed lines show the spin-orbit-split states with the Zeeman field $V_x=0$.\ The blue lines show that $V_x=0.1$ opens a gap with the width $2V_x$.\ This gap is a chemical potential window, when the chemical potential is in the gap, the Heterostructure exhibits only a single pair of Fermi points.\ Turning on a weak $\Delta$, the Majorana fermions emerge at two Fermi points when the condition of creating Majorana fermions $V_x>\Delta_{eff}^s$ is satisfied.\ Here the red lines show whether the condition of creating Majorana fermions $\Delta_{eff}^s$ is in the chemical potential window or not, the green and black dots represent the Fermi points. }
\label{fig1}
\end{figure}

As Fig.\ \ref{fig5} shows, the Heterostructure we consider is a semiconducting nanowire contacts with an ordinary superconductor.\ Because InAs is chosen to be the material of the semiconducting nanowire, a large spin-orbit coupling exists in the Heterostructure.\ Through the proximity effect between the semiconductor and superconductor, cooper pairs can leak into the nanowire and a $s$-wave pairing field emerges (which we  take to be real).\ A magnetic field $B_x$ is applied along $x$ direction, which produce a Zeeman splitting.\ Through varing the electron density of the nanowire periodically by applying the alternating gatevoltage $\widetilde{V}_{gate}$, the chemical potential of the Heterostructure changes periodically.\ The Hamiltonian of the Heterostructure reads $\mathcal{H=H_\mathrm{0}+H(\mathrm{t})}$, in which
\begin{shrinkeq}{-1.5ex}
\begin{equation}
\begin{split}
\mathcal{H}_0=\int &dx\psi^\dagger (-\frac{\partial_x^2}{2m}+V_x\sigma_x-i\alpha \partial_x\sigma_z )\psi\\
&+\Delta\int dx(\psi_\uparrow^\dagger\psi_\downarrow^\dagger+h. c),
\label{equ2}
\end{split}
\end{equation}
\end{shrinkeq}
\begin{shrinkeq}{-1ex}
\begin{equation}
\mathcal{H}(t)=-\int dx\psi^\dagger \mu(x,t)\psi,
\label{equ3}
\end{equation}
\end{shrinkeq}
where $x$ is the coordinate along the wire, $m$ is the effective mass of electron, $\sigma_{i}$ ($i=x,y,z$) are Pauli matrices.\ The operater $\psi_s$ annihilates an electron with spin $s=\uparrow\downarrow$.\ $\mu(x,t)=\mu(x,t+T)$ is a periodically driven chemical potential which is applied to the Heterostructure with period $T$.\ $V_x$ is Zeeman field which is offered by the magnetic field $B_x$, $\alpha$ is the strength of the spin-obit coupling, $\Delta$ is the real $s$-wave pairing field.\\
\indent In order to demonstrate that the Majorana fermions emerge when periodically driven chemical potential is applied.\ We first construct a lattice Hamiltonian that map onto continuum Hamiltonian $\mathcal{H}$ in the low density limit.\ This can be done in momentum space by replacing $p^{2}\rightarrow2({1}-\cos{p})$, $p\rightarrow \sin{p}$, $\int dp\rightarrow\frac{1}{L}\sum_{p}$, $\psi_p\rightarrow\sqrt{L}c_p$, $L$ is the Heterostructure size.\cite{Stoudenmire2011} Then we obtain Hamiltonian in momentum space $H=H_{0}(p)+H(p,t)$, where
\begin{shrinkeq}{-1ex}
\begin{equation}
\begin{aligned}
H_{0}(p)&=\frac{1}{L}\sum_{p}\frac{2(1-\cos{p})}{2m}c_p^\dagger c_p+V_x c_p^\dagger\sigma_x c_p\\
&+\alpha\sin{p}c_p^\dagger\sigma_z c_p+\Delta (c_{p\uparrow}^\dagger c_{-p\downarrow}^\dagger+h.c),\\
\label{equ32}
\end{aligned}
\end{equation}
\begin{equation}
H(p,t)=-\frac{1}{L}\sum_{p}\mu_{j}(t)c_{p}^\dagger c_{p},
\label{equ33}
\end{equation}
\end{shrinkeq}
where ${c_{p\alpha}^\dagger}$ ($ {c_{p\alpha}}$) creates (annihilates) an electron with momentum $p$ and spin $\alpha$ (up or down).\ We can replace $\sin{p}$ and $\cos{p}$ with the form $\sin{p}=(e^{ip}-e^{-ip})/{2i}$ and $\cos{p}=(e^{ip}+e^{-ip})/{2}$, then we transform the Hamiltonian to real space and obtain the lattice Hamiltonian $H=H_0+H(t)$,\cite{Stoudenmire2011} in which
\begin{shrinkeq}{-1ex}
\begin{equation}
\begin{aligned}
&\ \ \ \ \ \ \ \ \ H_0=\sum_{j}-J(c_{j}^\dagger c_{j+1}+c_{j}^\dagger c_{j-1})\\
&\ \ \ \ -\frac{i\alpha}{2}(c_{j}^\dagger\sigma_z c_{j+1}-c_{j}^\dagger\sigma_z c_{j-1})+{2}J c_{j}^\dagger c_{j}\\
&\ \ \ \ \ \ \ \ \ +V_x c_{j}^\dagger\sigma_x c_{j}+\Delta(c_{j\uparrow}^\dagger c_{j\downarrow}^\dagger+h. c),
\label{equ6}
\end{aligned}
\end{equation}
\begin{equation}
H(t)=-\sum_{j}\mu_{j}(t)c_{j}^\dagger c_{j},
\label{equ7}
\end{equation}
\end{shrinkeq}
where $ {c_{i\alpha}^\dagger}$ ($ {c_{i\alpha}}$) creates (annihilates) an electron on $ {i}$ site with spin $\alpha$ (up or down), $J=\hbar^2/{2ma^2}$ is the hopping strength and $\alpha_{latt}=\frac{\hbar}{a}\alpha_{cont}$ where $a$ is the lattice constant.\ Here it is convient to introduce the typical parameters with $m\sim0.05\ m_e$ where $m_e$ is the bare electron mass, $\alpha\sim0.1\ \mathrm{eV\mathring{A}}$, $V_x\sim1\ \mathrm{K}$ and $\Delta\sim1\ \mathrm{K}$, $J\sim1\ \mathrm{eV}$.\ These parameters suggest that the relevant hierarchy of energies $J \gg\alpha >\Delta$.\cite{Lutchyn2010,Stoudenmire2011} For simplify we set lattice constant $a=1$, $\hbar=1$, in the following we set $J=1$.\\
\indent In order to obtain the conditions of creating Majorana fermions in driven Heterostructure, it is convient to discuss the condition of creating Majorana fermions in static system.\ We change the Heterostructure to a static Heterostructure by setting $\mu(x,t)$ to a constant $\mu_0$.\cite{Kitaev2001,Lutchyn2010,Oreg2010} Considering the Zeeman field $V_x$ and the pairing field $\Delta$ vanish, the energy spectrum of the Hamiltonian is shown by the black dash lines in Fig.\ \ref{fig1}.\ For arbitrary values of $\mu_0$ above the minimum of the energy spectrum, the salient feature of these states is the generic presence of four Fermi points.\cite{Stoudenmire2011} \\
\indent Considering the situation with Zeeman field $V_x \not=0$, in this case, a gap is opened at $p=0$ as shown by the blue lines in Fig.\ \ref{fig1} where the width of the gap is $2V_x$.\ In this situation, the gap is a chemical potential window.\ When the chemical potential is in the gap, only two Fermi points exist and we can neglect the upper state of two states which is shown by blue lines.\ Turning on a weak $s$-wave pairing field $\Delta$, then two Majorana fermions appear at the left and right ends of the wire.\cite{Lutchyn2010,Stoudenmire2011} An analysis of the Bogoliubov-de Gennes equation reveals that the Majorana fermions exist only when the following condition $V_x >\Delta_{eff}^{s}$ is satisfied, where $\Delta_{eff}^{s}=\sqrt{\Delta^2+\mu_0^2}$ is the static effective pairing field.\ The condition is shown by the red lines in Fig.\ \ref{fig1}.\ When $V_x>\Delta_{eff}^s$, which is shown as the down red line, $\Delta_{eff}^s$ is in the chemical potential window and there are two Fermi points which are shown as the green dots, the Majorana fermions emerge at two Fermi points.\ When $V_x<\Delta_{eff}^s$, which is shown as the upper red line, $\Delta_{eff}^s$ is out of the chemical potential window and there are four Fermi points which are shown as the black dots, there are no Majorana fermions.\cite{Lutchyn2010} \\
\section{\label{sec3}QUASI-ENERGY SPECTRUM OF PERIODICALLY DRIVEN HETEROSTRUCTURE}
\subsection{Floquet's theory}
As the first step towards calculating the driven Heterostructure, we introduce the Floquet's theory briefly.\ When a Hamiltonian of the quantum system has a time-periodic dependence, i.e., $ {H(t)=H(t+T)}$ with ${T={2\pi}/{\omega}}$, the solution can be described by Floquet's theory.\cite{Grifoni1998} From Floquet's theory, we know that the Schr\"{o}dinger equation with a time-periodic dependent Hamiltonian has a complete set of solutions with the form $ {|\psi_n (t)\rangle=|u_n (t)\rangle exp({-i\varepsilon_n t}/{\hbar})}$.\ ${\varepsilon_n}$ is quasi-energy which characterizes the Floquet states in a system with the time translational symmetry $ {t\rightarrow t+T}$. The periodic function satisfies $ {|u_n(t)\rangle=|u_n(t+T)\rangle}$ with the eigenvalue equation
\begin{shrinkeq}{-1ex}
\begin{equation}
 {H_{eff}|u_n(t)\rangle=\varepsilon_n |u_n(t)\rangle},
\label{equ8}
\end{equation}
\end{shrinkeq}
where ${H_{eff}=H-i\hbar\partial_t}$  is the Floquet Hamiltonian.\ Note that the Floquet modes $ {|u_n (t)\rangle exp(im\omega t)}$ are also the solution of Eq.(\ref{equ8}), in which the shifted quasi-energy is ${\varepsilon_n+m\hbar\omega}$.\ ${\hbar\omega}$ is similar to the reciprocal lattice vector and we define the width of Brillouin zone with a sense of time.\ The integer $ {m=0,\pm 1,\pm 2\cdots}$ indexes the different Brillouin zones.\cite{liu2012,Eckardt2005} Because of the coupling between the spatial degree of freedom and temporal degree of freedom, it is convenient to introduce the Floquet basis\\
\begin{shrinkeq}{-2ex}
\begin{equation}
 {|\{n_i\},m\rangle=|\{n_i\}\rangle exp\{\frac{i}{\hbar}\int_{-\infty}^t dt'\sum_i \mu_{i}(t')n_i +im\omega t\}}.
\label{equ9}
\end{equation}
\end{shrinkeq}
$ {|\{n_i\}\rangle}$ indicates a Fock state with $ {n_i}$ particles on the $ {i}$th site, ${m}$ accounts for the Brillouin zones,\cite{Eckardt2005} $|\{n_i\},m\rangle$ consist of an extended Hilbert space of $ {T}$-periodic functions with the scalar product
\begin{shrinkeq}{-2ex}
\begin{equation}
 {\langle\langle\cdot|\cdot\rangle\rangle=\frac{1}{T}\int_0^T dt\langle\cdot|\cdot\rangle}.
\label{equ10}
\end{equation}
\end{shrinkeq}
The quasi-energies are obtained by computing the matrix elements of $ {H_{eff}}$ in the basis (\ref{equ9}) with respect to the scalar product (\ref{equ10}).\ The matrix elements in the Floquet Hamiltonian $ {H_{eff}}$ are
\begin{shrinkeq}{-1ex}
\begin{equation}
\begin{split}
&\ \ \ \ \ \ \ \ \ \ \ \ \ \ \ \ \ \ {\langle\langle\{{n_i}'\},m'|c_{i\alpha}^\dagger c_{j\beta}|\{n_i\},m\rangle\rangle=}\\
&{\frac{1}{T}\int_0^T dt\cdot exp\{i\int_{-\infty}^t dt' [\mu_j(t')-\mu_i(t')]-i(m'-m)\omega t\}},\\
&\ \ \ \ \ \ \ \ \ \ \ \ \ \ \ \ \ \ {\langle\langle\{{n_i}'\},m'|c_{i\alpha}^\dagger c_{j\beta}^\dagger|\{n_i\},m\rangle\rangle=}\\
& {\frac{1}{T}\int_0^T dt\cdot exp\{i\int_{-\infty}^t dt' [-\mu_i(t')-\mu_j(t')]-i(m'-m)\omega t\}},\\
&\ \ \ \ \ \ \ \ \ \ \ \ \ \ \ \ \ \ {\langle\langle\{{n_i}'\},m'|c_{i\alpha} c_{j\beta}|\{n_i\},m\rangle\rangle=}\\
&{\frac{1}{T}\int_0^T dt\cdot exp\{i\int_{-\infty}^t dt' [\mu_i(t')+\mu_j(t')]-i(m'-m)\omega t\}}.
\label{equ11}
\end{split}
\end{equation}
\end{shrinkeq}
In the above matrix, the diagonal block of the Floquet Hamiltonian $ {H_{eff}^{(m,m)}}$ is the m-Brillouin zone of quasi-energy, the nondiagonal blocks $ {H_{eff}^{(m',m)}}$ with $ {m'\not=m}$ corresponds to the interaction between different Brillouin zone.\cite{Inoue2010} When the driven potential $\mu(x,t)$ is relatively small and the adiabatic condition $ {J\ll \hbar\omega}$ is satisfied, the interactions between different Brillouin zone is negligible, in this case, the driven system behaves similar to the static system with $\mu(x,t)=\mu_0$.\cite{Inoue2010} Now, suppose that we enhance the driven potential $\mu(x,t)$ or reduce the driven frequency, then we have to consider the coupling of different Brillouin zones. \\
\indent Let's consider the simplest form of the space-independent driven chemical potential
\begin{shrinkeq}{-1ex}
\begin{equation}
 {\mu(x,t)=\mu(t)=\mu+\mu\cos{\omega t}},
\label{equ12}
\end{equation}
\end{shrinkeq}
where $\mu$ is the driven amplitude and $\omega$ is the driven frequency.\ From Eq. (\ref{equ12}) we obtain
\begin{center}
\begin{shrinkeq}{-5ex}
\begin{equation}
\begin{split}
&\ \ \ \ \ \ \ \ \ \ \ \ \ \langle\langle\{{n_i}'\},m'|c_{i\alpha}^\dagger c_{j\beta}|\{n_i\},m\rangle\rangle=\delta_{m'm},\\
&\ \ \ \ \ \ \ \ \ \ \ \ \ \ \ \ \langle\langle\{{n_i}'\},m'|c_{i\alpha}^\dagger c_{j\beta}^\dagger|\{n_i\},m\rangle\rangle=\\
&\ \ \ \ {\frac{1}{T}\int_0^T dt\cdot exp\{-i\frac{2\mu}{\omega}\sin{\omega t}-i(m'-m+\frac{2\mu}{\omega})\omega t\}},\\
&\ \ \ \ \ \ \ \ \ \ \ \ \ \ \ \ \langle\langle\{{n_i}'\},m'|c_{i\alpha} c_{j\beta}|\{n_i\},m\rangle\rangle=\\
&\ \ \ {\frac{1}{T}\int_0^T dt\cdot exp\{i\frac{2\mu}{\omega}\sin{\omega t}-i(m'-m-\frac{2\mu}{\omega})\omega t\}},
\label{equ13}
\end{split}
\end{equation}
\end{shrinkeq}
\end{center}
where the integrals of \eqref{equ13} can be viewed as a function of ${2\mu}/{\omega}$.\ When ${2\mu}/{\omega}$ is an integer, the integrals of \eqref{equ13} are Bessel functions of integer order.\ From the form of integrals \eqref{equ13}, the values of $H_{eff}^{(m',m)}$ depend on $m'-m$.\ The diagonal blocks $H_{eff}^{(m,m)}$ and nondiagonal blocks $H_{eff}^{(m',m)}$ have the form of $H_{eff}^{(0)}+m\hbar\omega$ and $H_{eff}^{(m'-m)}$, respectively.\\
\subsection{The quasi-energy spectrum of neglecting the interactions between different Brillouin zones}
Considering ${2\mu}/{\omega}$ is an integer first.\ In this case, the integrals of (\ref{equ13}) are Bessel functions of integer order, then ${H_{eff}}$ in real space can be expressed as
\begin{figure*}
\subfigure{\includegraphics[scale=0.3]{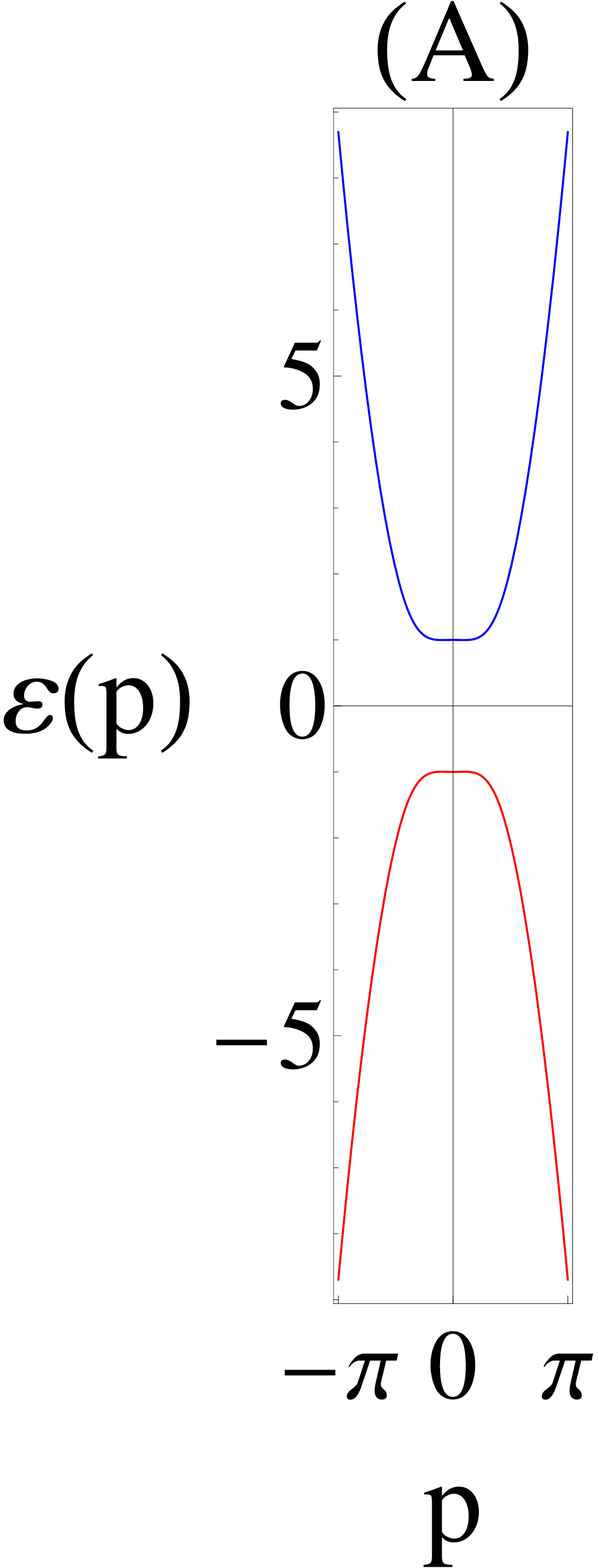}}
\subfigure{\includegraphics[scale=0.3]{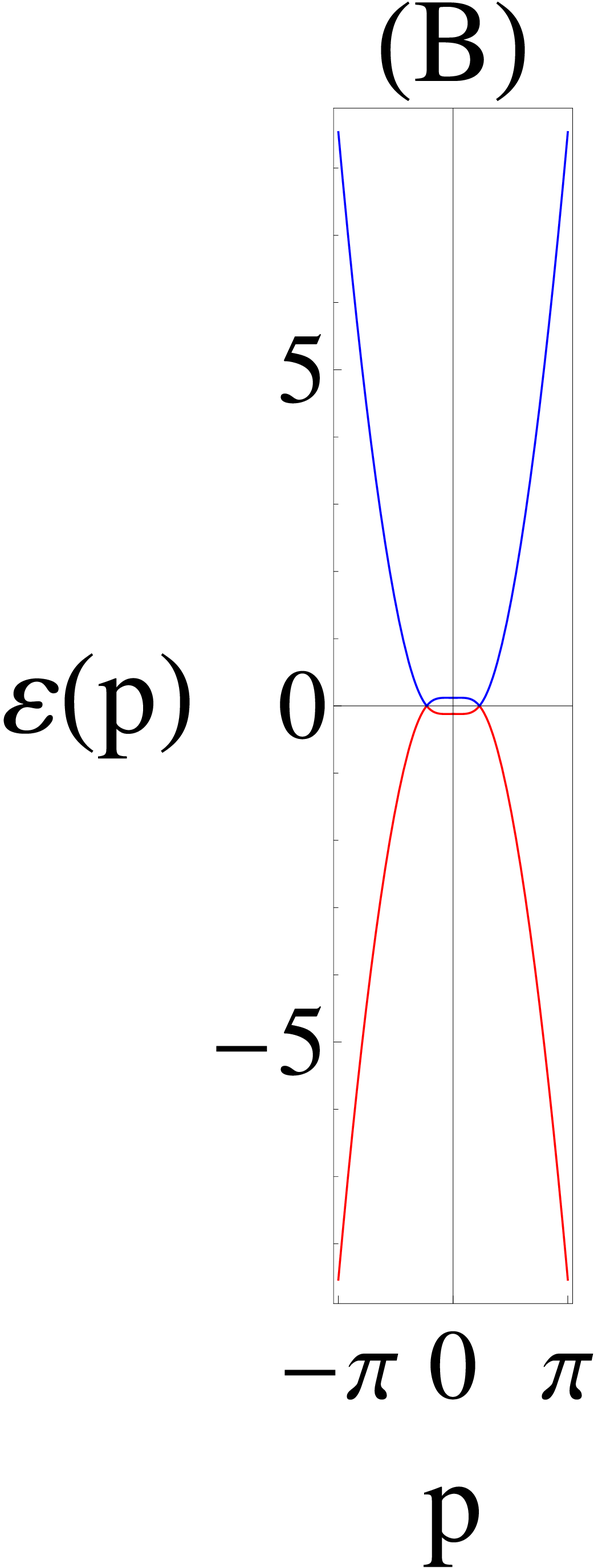}}
\subfigure{\includegraphics[scale=0.3]{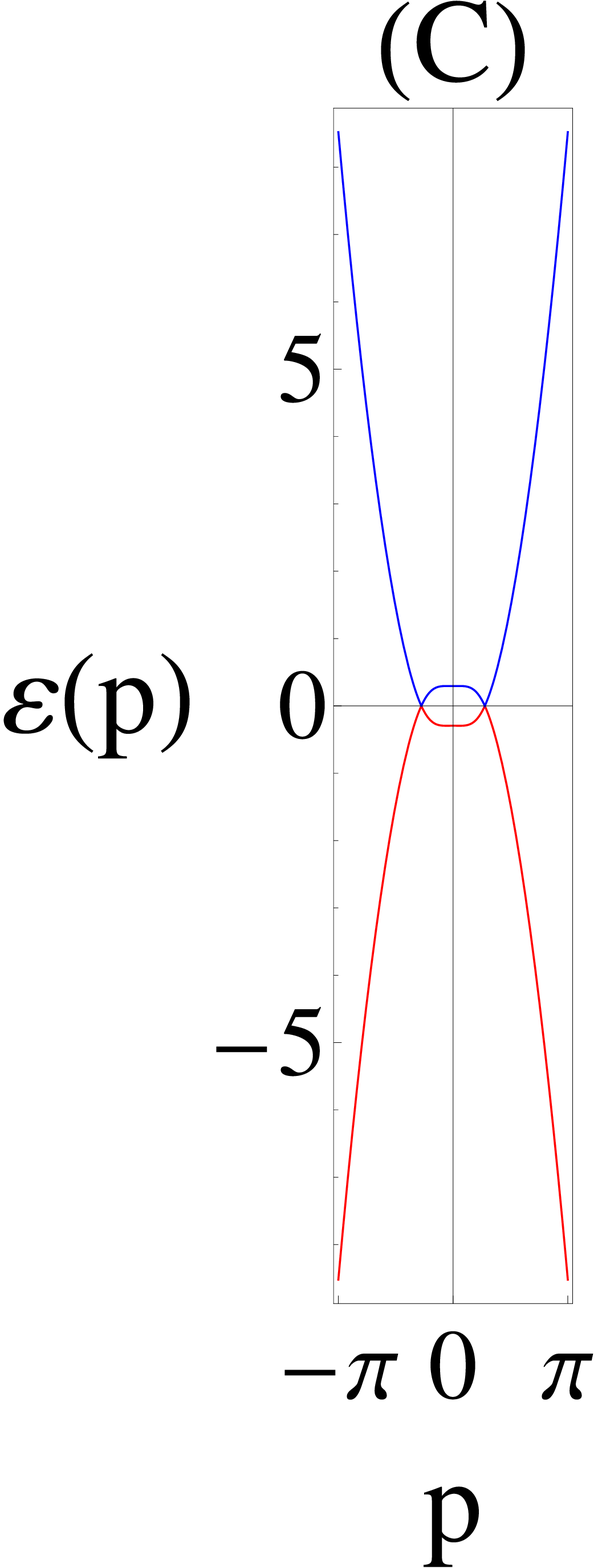}}
\subfigure{\includegraphics[scale=0.3]{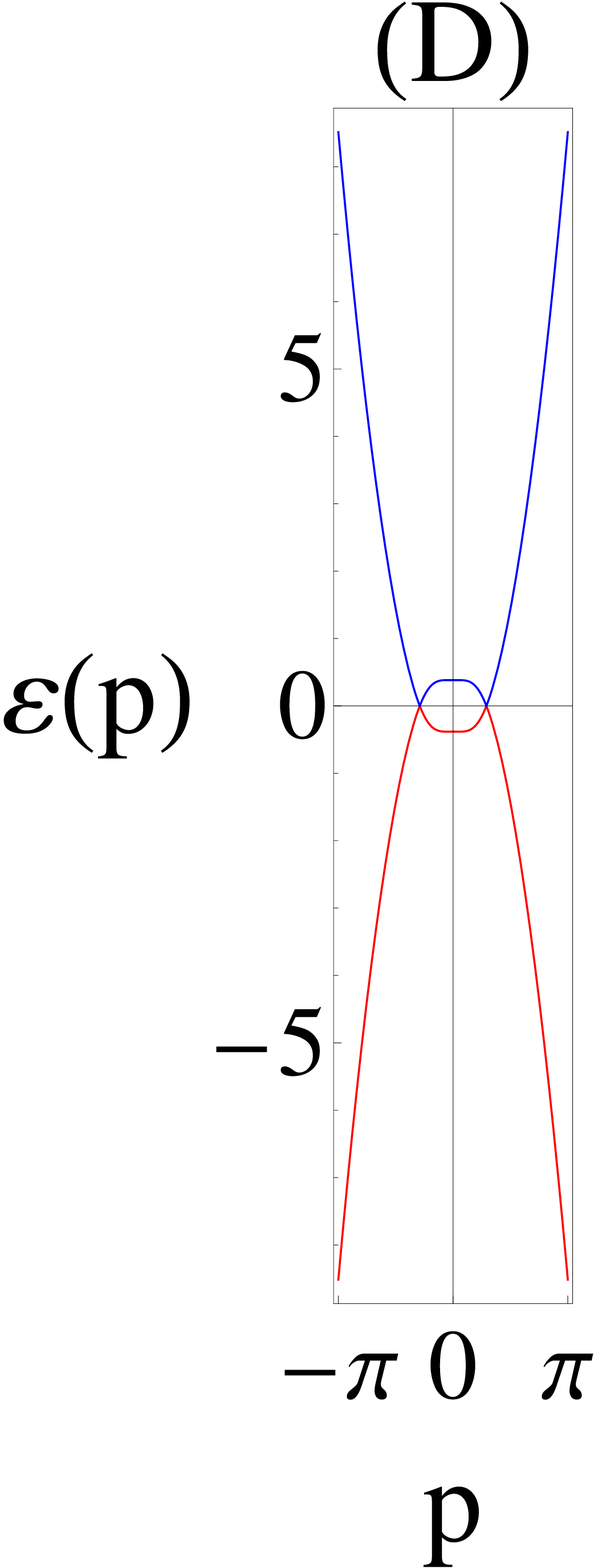}}
\caption{Quasi-energy spectrums $\varepsilon(p)$ of the Hamiltonian ${H_{eff}}$ in Eq.\ \eqref{equ23} when $\Delta=2$, $V_x=1$ (setting $J$ to be unit).\ (A) shows ${2\mu}/{\omega}=0$ with $\mu\rightarrow0$, there are no Majorana fermions.\ The Majorana fermions emerge by turning up the amplitude $\mu$ continually as (B) with ${{2\mu}/{\omega}=1}$, (C) with $ {{2\mu}/{\omega}=2}$ and (D) with $ {{2\mu}/{\omega}=3}$.\ The momentum is restricted to the first Brillouin zone and just lowest states is plotted. }
\label{fig2}
\end{figure*}
\begin{shrinkeq}{-2ex}
\begin{equation}
\begin{aligned}
H_{eff}^{(0)}&=\sum_{j}-J(c_{j}^\dagger c_{j+1}+c_{j}^\dagger c_{j-1})+V_x c_{j}^\dagger\sigma_x c_{j}\\
&{-\frac{i\alpha}{2}(c_{j}^\dagger\sigma_z c_{j+1}-c_{j}^\dagger\sigma_z c_{j-1})+{2}J c_{j}^\dagger c_{j}}\\
&{+J_{\frac{2\mu}{\omega}}(-\frac{2\mu}{\omega})\Delta c_{j\uparrow}^\dagger c_{j\downarrow}^\dagger+J_{-\frac{2\mu}{\omega}}(\frac{2\mu}{\omega})\Delta c_{j\downarrow}c_{j\uparrow}},
\label{equ18}
\end{aligned}
\end{equation}
\end{shrinkeq}
\begin{shrinkeq}{-2ex}
\begin{equation}
\begin{aligned}
H_{eff}^{(n)}=\sum_{j}J_{n+\frac{2\mu}{\omega}}(-\frac{2\mu}{\omega})\Delta c_{j\uparrow}^\dagger c_{j\downarrow}^\dagger+J_{n-\frac{2\mu}{\omega}}(\frac{2\mu}{\omega})\Delta c_{j\downarrow} c_{j\uparrow},
\label{equ19}
\end{aligned}
\end{equation}
\end{shrinkeq}
where ${J_{n}}$ is the Bessel function of $ {n}$th order and $ {n=m'-m}$.\ In momentum space, the explicit expression of the effect Hamiltonian is
\begin{widetext}
\begin{shrinkeq}{-2ex}
\begin{equation}
H_{eff}=\left(\begin{array}{ccccc}
\ddots& \vdots& \vdots& \vdots& \quad \\
\ldots& {H_p^{(0)}}& {H_p^{(1)}}& {H_p^{(2)}}&\ldots\\
\ldots& {H_p^{(-1)}}& {H_p^{(0)}+\omega}& {H_p^{(1)}}&\ldots\\
\ldots& {H_p^{(-2)}}& {H_p^{(-1)}}& {H_p^{(0)}+2\omega}&\ldots\\
\quad &\vdots&\vdots&\vdots&\ddots
\end{array}\right),
\label{equ20}
\end{equation}
\end{shrinkeq}
\begin{equation}
H_p^{(0)}=\left(\begin{array}{cccc}
{\frac{p^2}{2m}}-\mu_0+\alpha p&V_x&0&J_{\frac{2\mu}{\omega}}(-\frac{2\mu}{\omega})\Delta\\
V_x&\frac{p^2}{2m}-\mu_0-\alpha p&-J_{\frac{2\mu}{\omega}}(-\frac{2\mu}{\omega})\Delta&0\\
0&-J_{-\frac{2\mu}{\omega}}(\frac{2\mu}{\omega})\Delta&-\frac{p^2}{2m}+\mu_0-\alpha p&-V_x\\
J_{\frac{-2\mu}{\omega}}(\frac{2\mu}{\omega})\Delta&0&-V_x&-\frac{p^2}{2m}+\mu_0+\alpha p\\
\end{array}\right),
\label{equ21}
\end{equation}
\begin{equation}
H_p^{(n)}=\left(\begin{array}{cccc}
0&0&0&J_{n+\frac{2\mu}{\omega}}(-\frac{2\mu}{\omega})\Delta\\
0&0&-J_{n+\frac{2\mu}{\omega}}(-\frac{2\mu}{\omega})\Delta&0\\
0&-J_{n-\frac{2\mu}{\omega}}(\frac{2\mu}{\omega})\Delta&0&0\\
J_{n-\frac{2\mu}{\omega}}(\frac{2\mu}{\omega})\Delta&0&0&0\\
\end{array}\right).
\label{equ22}
\end{equation}
\end{widetext}
\indent Assuming $\omega$ is sufficient large so that the interactions between different Brillouin zones can be neglected, the effective Hamiltonian becomes
\begin{equation}
H_{eff}=\left(\begin{array}{ccccc}
\ddots& \vdots& \vdots& \vdots& \quad \\
\ldots& {H_p^{(0)}}& 0& 0&\ldots\\
\ldots& 0& {H_p^{(0)}+\omega}&0&\ldots\\
\ldots& 0& 0& {H_p^{(0)}+2\omega}&\ldots\\
\quad &\vdots&\vdots&\vdots&\ddots
\end{array}\right).
\label{equ23}
\end{equation}
\indent We choose the driven frequency $\omega=20$ and diagonalize the matrix of $H_{eff}$ directly to obtain the quasi-energy spectrums which are shown by Fig.\ \ref{fig2}.\ In this case, the driven Heterostructure is the same as the static Heterostructure, but the condition of creating Majorana fermions is replaced by
\begin{equation}
V_x>|\Delta_{eff}^{d}|,
\label{equ4}
\end{equation}
where the effective pairing field has the form of ${\Delta_{eff}^{d}=J_{\pm{2\mu}/{\omega}}(\mp{2\mu}/{\omega})\Delta}$.\ ${\Delta_{eff}^{d}}$ decreases when ${{2\mu}/{\omega}}$ increases.\ When the condition \eqref{equ4} is satisfied, the Majorana fermions emerge, which are shown by Fig.\ \ref{fig2} (B), Fig.\ \ref{fig2} (C) and Fig.\ \ref{fig2} (D). \\
\indent Now let's consider ${2\mu}/{\omega}$ is a real number.\ In this situation, the values of the integrals \eqref{equ13} are no longer real number but complex number.\ Because of the conjugate of the last two integrals in \eqref{equ13}, the Hamiltonian $H_{eff}$ is still a Hermitian operator, which in real space can be expressed as
\begin{shrinkeq}{-1ex}
\begin{equation}
\begin{aligned}
&H_{eff}^{(0)}=\sum_{j}-J(c_{j}^\dagger c_{j+1}+c_{j}^\dagger c_{j-1})+V_x c_{j}^\dagger\sigma_x c_{j}\\
&\ \ \ \ -\frac{i\alpha}{2}(c_{j}^\dagger\sigma_z c_{j+1}-c_{j}^\dagger\sigma_z c_{j-1})+{2}J c_{j}^\dagger c_{j}\\
&\ +J_{\frac{2\mu}{\omega}}'(-\frac{2\mu}{\omega})\Delta c_{j\uparrow}^\dagger c_{j\downarrow}^\dagger+J_{-\frac{2\mu}{\omega}}'(\frac{2\mu}{\omega})\Delta c_{j\downarrow}c_{j\uparrow},
\label{equ24}
\end{aligned}
\end{equation}
\begin{equation}
\begin{aligned}
H_{eff}^{(n)}&=\sum_{j}J_{n+\frac{2\mu}{\omega}}'(-\frac{2\mu}{\omega})\Delta c_{j\uparrow}^\dagger c_{j\downarrow}^\dagger\\
&+J_{n-\frac{2\mu}{\omega}}'(\frac{2\mu}{\omega})\Delta c_{j\downarrow} c_{j\uparrow},
\label{equ25}
\end{aligned}
\end{equation}
\end{shrinkeq}
where the functions $J_{n-{2\mu}/{\omega}}'({2\mu}/{\omega})$ and $J_{n+{2\mu}/{\omega}}'(-{2\mu}/{\omega})$ are
\begin{shrinkeq}{-1ex}
\begin{equation}
\begin{aligned}
&\ \ \ \ \ \ \ \ \ \ \ \ \ \ \ \ \ \ \ \ \ J_{n-\frac{2\mu}{\omega}}'(\frac{2\mu}{\omega})=\\
&{\frac{1}{T}\int_0^T dt\cdot exp\{i\frac{2\mu}{\omega}\sin{\omega t}-i(m'-m-\frac{2\mu}{\omega})\omega t\}},
\label{equ26}
\end{aligned}
\end{equation}
\begin{equation}
\begin{aligned}
&\ \ \ \ \ \ \ \ \ \ \ \ \ \ \ \ \ \ \ \ \ J_{n+\frac{2\mu}{\omega}}'(-\frac{2\mu}{\omega})=\\
&{\frac{1}{T}\int_0^T dt\cdot exp\{-i\frac{2\mu}{\omega}\sin{\omega t}-i(m'-m+\frac{2\mu}{\omega})\omega t\}}.
\label{equ27}
\end{aligned}
\end{equation}
\end{shrinkeq}
We neglect $H^{(n)}$ by choosing a large $\omega$.\ The effective Hamiltonian $H_{eff}$ in momentum space has the form of Eq.\ \eqref{equ23}, in which $H_p^{(0)}$ has the form of Eq.\ \eqref{equ21} but the Bessel function $J_{-{2\mu}/{\omega}}$ and $J_{{2\mu}/{\omega}}$ are replaced by $J_{-{2\mu}/{\omega}}'$ and $J_{{2\mu}/{\omega}}'$.\\
\begin{figure*}
\includegraphics[scale=0.2]{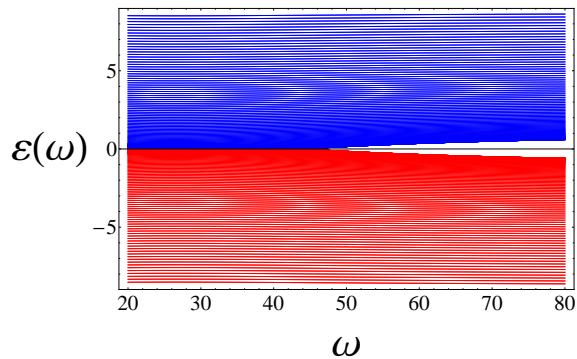}
\caption{Quasi-energy spectrums $\varepsilon(\omega)$ of the Hamiltonian $H_{eff}$ in Eq.\ \eqref{equ23} when $\Delta=2$, $V_x=1$, $\mu=20$ (setting $J$ as unit), the frequency $\omega$ is tuned from $80$ to $20$ continually (setting $J=1$ to be unit).\ Turning down the frequency of the driven potential $ {\omega}$ continually, the Majorana fermions emerge.\ The momentum is restricted to the first Brillouin zone and just lowest states is plotted.}
\label{fig3}
\end{figure*}
\begin{figure*}
\subfigure{\includegraphics[scale=0.3]{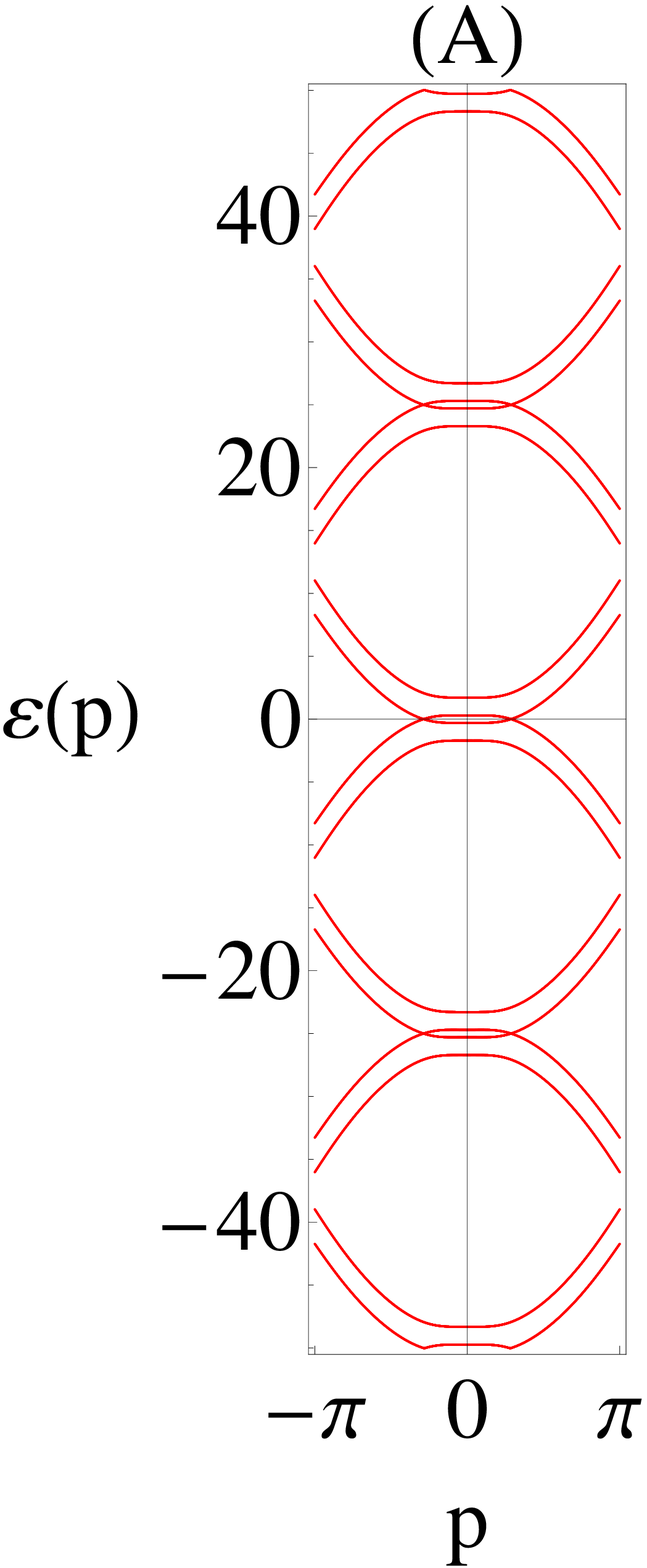}}
\subfigure{\includegraphics[scale=0.3]{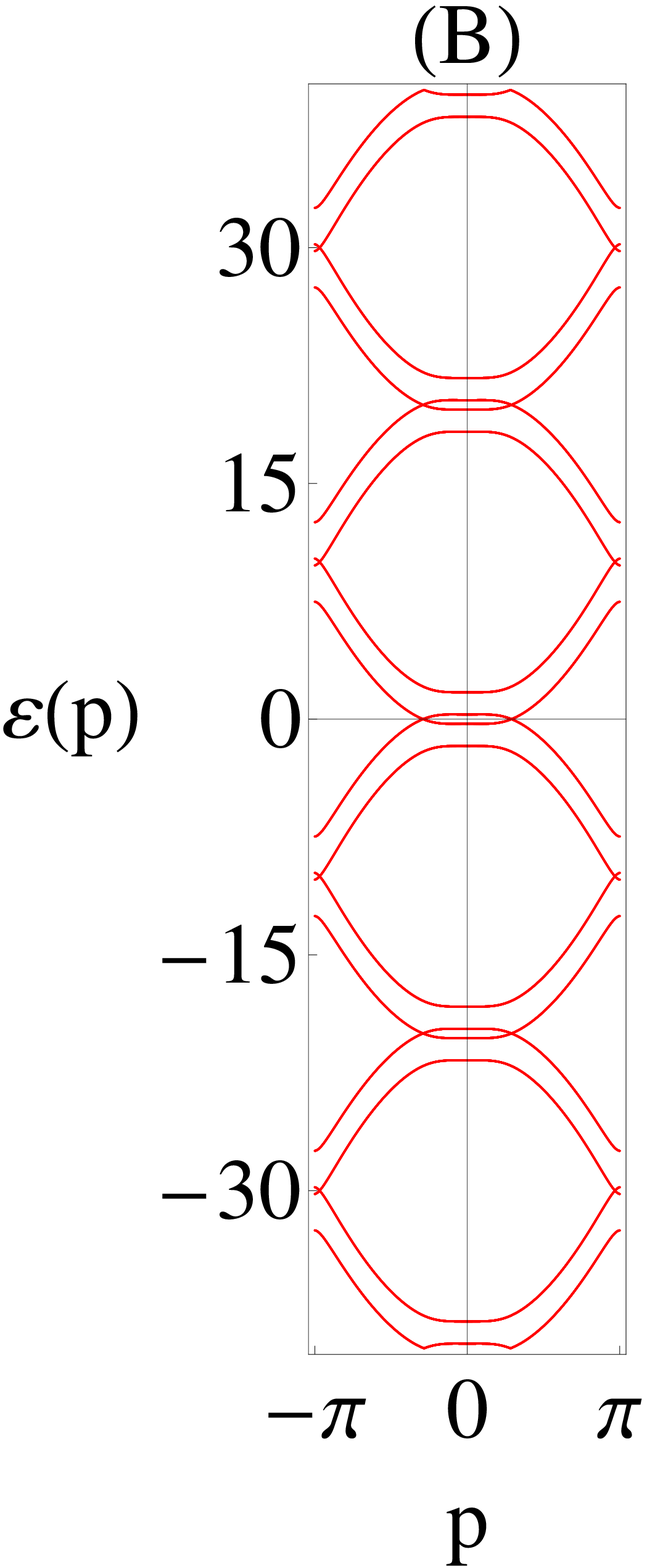}}
\subfigure{\includegraphics[scale=0.3]{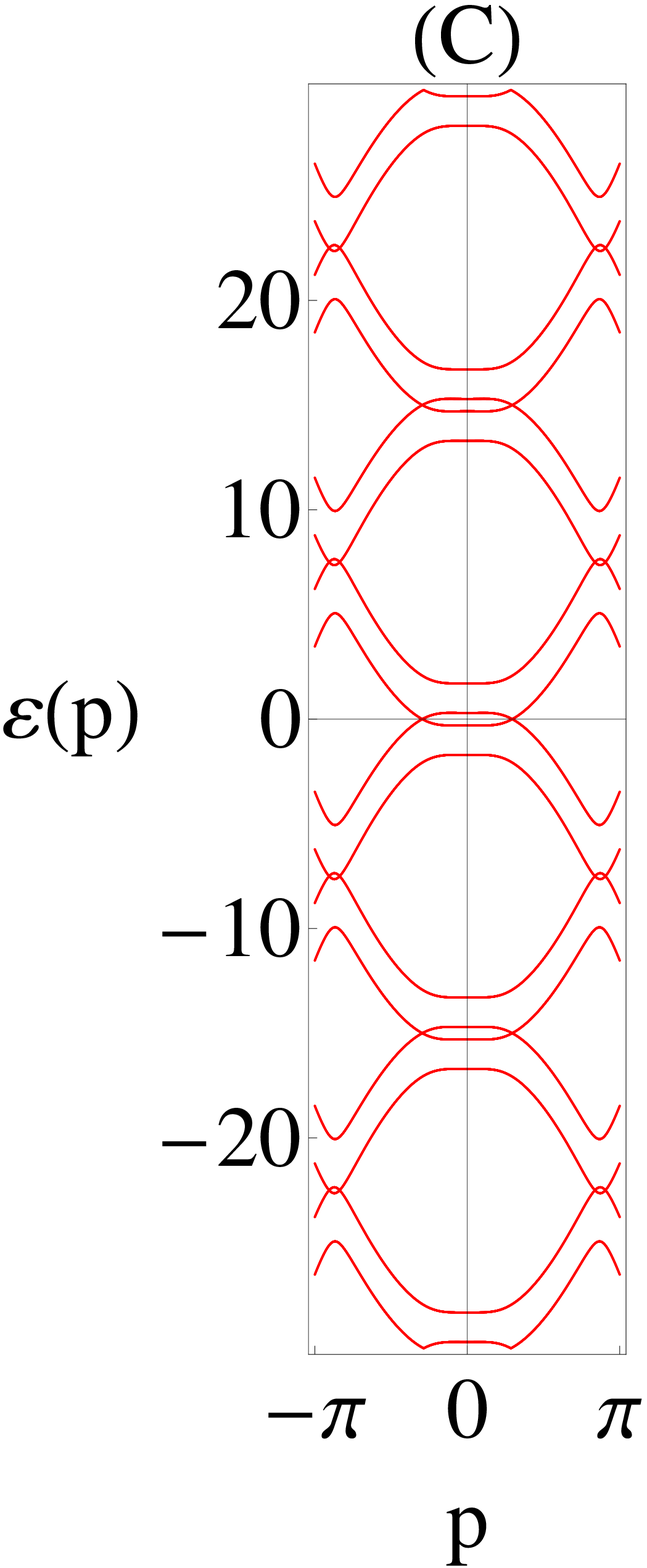}}
\subfigure{\includegraphics[scale=0.3]{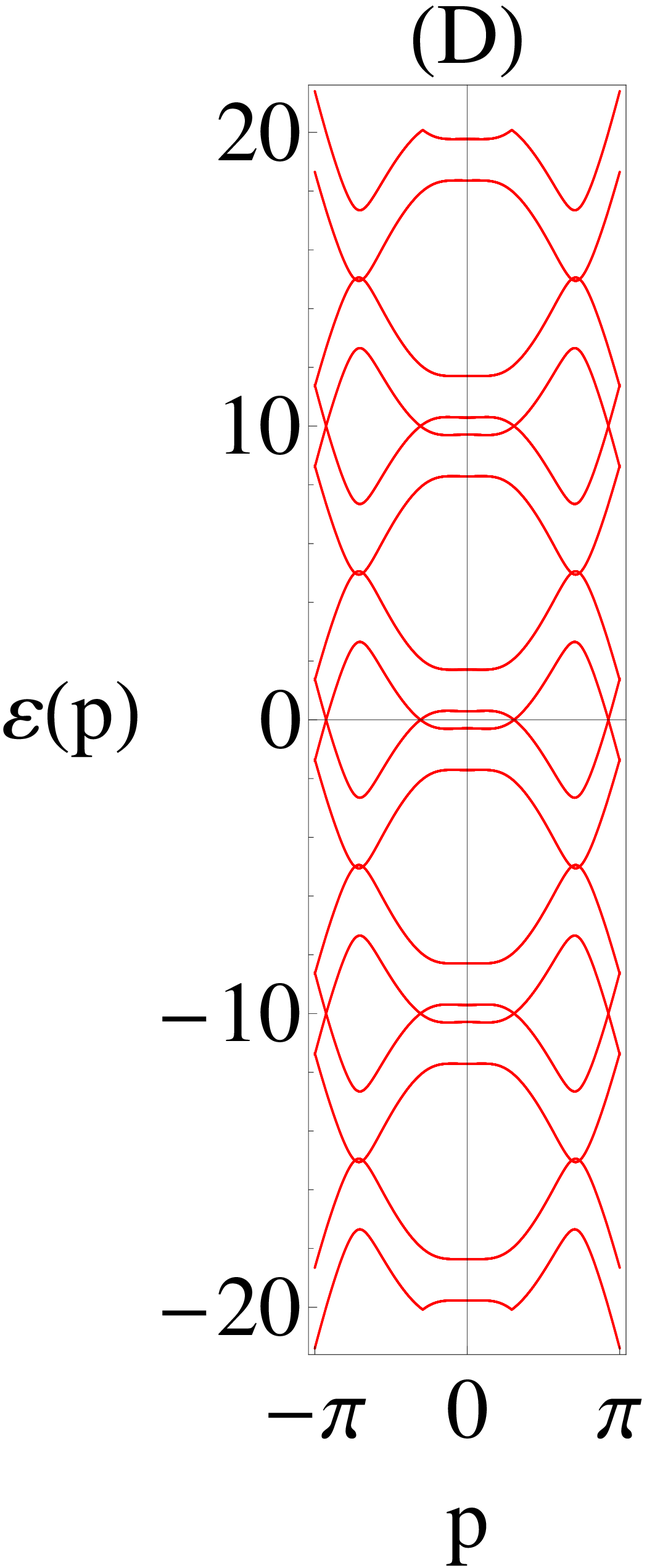}}
\caption{Quasi-energy spectrums of the Hamiltonian $ {H_{eff}}$ in Eq.\ \eqref{equ31} when $\Delta=2$, $V_x=1$, ${2\mu}/{\omega}=2$ (setting $J$ as unit).\ The spectrums are showed with (A) as $\omega=25$, (B) as $\omega=20$, (C) as $\omega=15$ and (D) as $\omega=10$.\ From (A) to (C), the interaction between the nearest Brillouin zones of quasi-energy is becoming stronger, but there are still two fermi points as no interaction case, the Majorana fermions exist.\ When the $\omega=10$ as (D), there are four fermi point emerge, there are no Majorana fermions.\ The momentum is restricted to the first Brillouin zone.}
\label{fig4}
\end{figure*}
\indent We choose $\Delta=2$, $V_x=1$, $\mu=20$ and diagonalize $H_{eff}$ directly.\ We tune the frequency $\omega$ continuously from $80$ to $20$, in this case, ${2\mu}/{\omega}$ change from $0.5$ to $2$.\ The quasi-energy spectrums are shown as Fig.\ \ref{fig3}.\ From Fig\ \ref{fig3} we can find that the Majorana fermions emerge when the frequency $\omega$ is below a critical frequency $\omega$ (about 45).\ Here the driven effective pairing field has the form of
\begin{shrinkeq}{-1ex}
\begin{equation}
|\Delta_{eff}^{d}|^2=J_{-\frac{2\mu}{\omega}}'(\frac{2\mu}{\omega})\Delta\cdot J_{\frac{2\mu}{\omega}}'(-\frac{2\mu}{\omega})\Delta,
\label{equ30}
\end{equation}
\end{shrinkeq}
with the frequency decreasing, $|\Delta_{eff}|^2$ also decreases.\ On the other hand, utilising Eq.\ \eqref{equ30} and the condition \eqref{equ4}, the critical frequency $\omega$ which induces the Majorana fermions is approach $45$, ${2\mu}/{\omega}$ is approach $0.9$.\ The result agree with the Fig.\ \ref{fig3} where the Majorana fermions emerge.
\subsection{The quasi-energy spectrum of considering the interactions between nearest Brillouin zones}
When the frequency $\omega$ is not sufficient large, the interactions between different Brillouin zones become significant.\ Now we take the interactions between nearest Brillouin zones into account and neglect the other parts of $H_{eff}$.\ Then $H_{eff}$ can be written as\\
\begin{shrinkeq}{-2ex}
\begin{equation}
H_{eff}=\left(\begin{array}{ccccc}
\ddots& \vdots& \vdots& \vdots& \quad \\
\ldots& {H_p^{(0)}}& H_p^{(1)}& 0&\ldots\\
\ldots& H_p^{(-1)}& {H_p^{(0)}+\omega}&H_p^{(1)}&\ldots\\
\ldots& 0& H_p^{(-1)}& {H_p^{(0)}+2\omega}&\ldots\\
\quad &\vdots&\vdots&\vdots&\ddots
\end{array}\right),
\label{equ31}
\end{equation}
\end{shrinkeq}
where $H_p^{(0)}$, $H_p^{(-1)}$ and $H_p^{(1)}$ have the forms of Eq. \eqref{equ21}, Eq. \eqref{equ22}.\ We choose ${2\mu}/{\omega}=2$ in $H_p^{(0)}$, $H_p^{(-1)}$ and $H_p^{(1)}$, the case of ${2\mu}/{\omega}=2$ has been discussed above in no interaction case, in which the Majorana fermions emerge.\ With diagonalizing the matrix of $H_{eff}$ in Fig.\ \eqref{equ31}, the numerical results are shown in Fig.\ \ref{fig4}.\ In Fig.\ \ref{fig4} (A), the interactions between nearest Brillouin zones are very weak and the quasi-energy spectrum is similar to no interaction case, there are just two Fermi point and the Majorana fermions exist.\ When $\omega$ decreases as Fig.\ \ref{fig4} (B) and Fig.\ \ref{fig4} (C), the interactions become stronger, the quasi-energy spectrums near momentum $\pm\pi$ change rapidly, but the center part of the energy spectrums are still the same as no interaction case, the Majorana fermions still exist.\ Finally, as Fig.\ \ref{fig4} (D) shows, when $\omega$ keeps decreasing, the other two Fermi points emerge and two Majorana fermions disappear.\cite{Lutchyn2010}\ With the center part of spectrum changing, the calculation bases on no interaction case becomes totally invalid.
\section{\label{sec4}EXPERIMENTAL PROTOCOL}
As Fig.\ \ref{fig5} shows, an semiconducting nanowire is arranged to contact with an $s$-wave superconductor.\ The superconductor is separated from the Si substrate by a SiO$_2$ layer.\cite{Doh2005,van2006} Through the proximity effect the Cooper pairs from a superconductor leak into the nanowire.\ Due to the weak capacitive coupling between the nanowire and the Si substrate, we can apply an alternating gatevoltage $\widetilde{V}_{gate}$ to the Si substrate to vary the electron density in the nanowire, which changes the chemical potential of the nanowire periodically.\cite{Doh2005} The magnetic field $B_x$ is applied to open a gap at zero momentum and eliminate fermion doubling.\\
\indent Here we choose InAs and Nb or Al to be the materials of the nanowire and $s$-wave superconductor, respectively, which can form a highly transparent interface for electrons between the nanowire and the superconductor.\ Moreover, because of the different Lande factors $\mathrm{g}$ with $\mathrm{g}_{InAs} \le 35$ and $\mathrm{g}_{Nb} \sim 1$, we can apply a suitable $B_x$ to open a sizable gap without destroying the superconduction in superconductor.\cite{Lutchyn2010}\\
\indent The InAs nanwire is grown via a catalytic based on a vapor-liquid-solid mechanism with diameters ranging from $40$ to $130\ \mathrm{nm}$  and lengths of $3$ to $10\ \mathrm{\mu m}$.\cite{Doh2005} The temperature of the Heterostructure should be low enough so that the de-broglie wavelength of electron can be long enough, the Heterostructure can be viewed as a 1D Heterostructure.\ In practice, the temperature of Heterostructure is of the order of $\sim \mathrm{mK}$.\\
\indent In order to create Majorana fermions in such a driven Heterostructure, it is necessary to obtain the rough scales of experimental parameters.\ From the typical parameters we obtain the magnetic field $B_x$ is less than $0.1\ \mathrm{T}$,\cite{Lutchyn2010} the driven amplitude $\mu$ is of the order of $10\ \mathrm{K}$ and the driven frequency $\omega$ is about $10^{12}\sim10^{13}\ \hertz$.\\
\section{\label{sec5}CONCLUSION}
In summary, we propose a new approach to create Majorana fermions in a periodically driven semiconductor-superconductor Heterostructure.\ By using Floquet's theory, we calculate the quasi-energy spectrums of the case which neglect the interaction between different Brillouin zones of quasi-energy.\ Then we demonstrate when the pairing field $\Delta$ and Zeeman splitting $V_x$ have certain values, Majorana fermions can be created under following condition with ${2\mu}/{\omega}$ being larger than a minimum to make sure the effective pairing field $\Delta_{eff}^{d}<V_x$.\ Furthermore, By calculating the case of considering the nearest Brillouin zones of quasi-energy, the condition of creating Majorana fermions is restricted in which the driven frequency $\omega$ should be higher than a lower limit.\ Here the large driven frequency avoid the interactions between different Brillouin zones being too strong.\ For example, when the hopping strength $J$ is chosen to be the unit, $\Delta=2$ and $V_x=1$, in order to create Majorana fermions in the Heterostructure, the conditions which ${2\mu}/{\omega}>0.9$ and $\omega>10$ should be satisfied to make sure $\Delta_{eff}^{d}<V_x$ and the interaction between different Brillouin zones of quasi-energy being weak enough .\ Finally, we discuss an experimental proposal of creating Majorana fermions.\ We hope our work will be useful to the future experimental detection of Majorana fermions.
\begin{acknowledgments}
We are grateful to Guocai Liu and Yuren Shi for helpful discussions. This work was supported by the NKBRSFC under grants Nos. 2011CB921502, 2012CB821305, 2009CB930701, 2010CB922904, NSFC under grants Nos. 10934010, 11228409, 61227902,11065010 and NSFC-RGC under grants Nos. 11061160490 and 1386-N-HKU748/10.
\end{acknowledgments}

\end{document}